\documentclass[aps,prl,superscriptaddress,showpacs,reprint]{revtex4-1}

\usepackage{color}
\usepackage{natbib}
\usepackage[pdftex]{graphicx}
\usepackage{epstopdf}
\usepackage[latin1]{inputenc}
\usepackage[ngerman,english]{babel} 
\begin{document}

\title{Core Hole-Electron Correlation in Coherently Coupled Molecules}

\author{M. Scholz}
\author{F. Holch}
\author{C. Sauer}
\author{M. Wiessner}
\author{A. Sch\"oll}
\email[corresponding author. Email: ]{achim.schoell@physik.uni-wuerzburg.de}
\author{F. Reinert}
\affiliation{Experimentelle Physik VII and R\"ontgen Research Center for Complex Material Systems RCCM, Universit\"at W\"urzburg, 97074 W\"urzburg, Germany}
\affiliation{Gemeinschaftslabor f\"ur Nanoanalytik, Karlsruher Institut f\"ur Technologie KIT, 76021 Karlsruhe, Germany}

\date{\today}

\begin{abstract}
We study the core hole-electron correlation in coherently coupled molecules by energy dispersive near edge x-ray absorption fine-structure spectroscopy. In a transient phase, which exists during the transition between two bulk arrangements, 1,4,5,8-naphthalene-tetracarboxylicacid-dianhydride multilayer films exhibit peculiar changes of the line shape and energy position of the x-ray absorption signal at the C \textit{K}-edge with respect to the bulk and gas phase spectra. By a comparison to a theoretical model based on a coupling of transition dipoles, which is established for optical absorption, we demonstrate that the observed spectroscopic differences can be explained by an intermolecular delocalized core hole-electron pair. By applying this model we can furthermore quantify the coherence length of the delocalized core exciton.
\end{abstract}

\maketitle 


X-ray absorption measurements at high-resolution beam lines of third-generation synchrotron sources are a powerful tool for the investigation of the electronic and vibronic structure of molecules \cite{hofer_photoemission_1990, fohlisch_beyond_1998, scholl_electron-vibron_2004}. From an analysis of high-resolution data one can gain insight into the complex response  of the electronic system and of the molecular frame on the electronic excitation. 
Upon excitation of a core electron into an unoccupied molecular orbital, a particularly strong reaction of the electronic system in the vicinity of the core excitation occurs. This core hole-electron correlation, which is often referred to as core exciton, has to be included for a correct description of the experimental data. However, the nature of core excitons is not clear in molecular materials, since experimental information, e.g., on the spatial expansion, is usually complicated to extract from x-ray absorption data due to various other interfering effects \cite{morar_observation_1985, bruhwiler_^*_1995,zou_solid_2006}. Thus, it is not clear whether the excitation is mainly localized on one molecule or on a subgroup of a molecule, or if it extends over several molecules. 

In this work we demonstrate the importance of the core hole-electron correlation in near edge x-ray absorption fine-structure (NEXAFS) data of organic molecules. On the example of the organic dye 1,4,5,8-naphthalene-tetracarboxylicacid-dianhydride (NTCDA) we show that under certain circumstances the molecules exhibit an extraordinarily strong intermolecular coupling, which leads to a redshift of electronic transitions and a substantial narrowing of the vibronic progressions in the NEXAFS spectra with respect to the gas phase. The respective experimental data can be described by a coupling of the transition dipoles to the neighboring molecules, thus resulting in a significant intermolecular delocalization of the core hole-electron pair. While this effect is well established for optical excitations \cite{muller_exciton_2011,wurthner_j-aggregates:_2011}, the calculated transition dipoles in core excitations are substantially smaller and thus a more local character of the excitation may generally be expected. 

The delocalization of the core excition upon x-ray absorption implies not only the transfer of the excited electron but also of the core hole. While for diatomic molecules, such as $N_2$ and $O_2$, experimental evidence for the nonlocal character of the core hole exists \cite{bjorneholm_doppler_2000, rolles_isotope-induced_2005, schoffler_ultrafast_2008}, an intermolecular delocalization of a core hole is not easily understood. 

\begin{figure}
	\centering
			\includegraphics[width=0.50\textwidth]{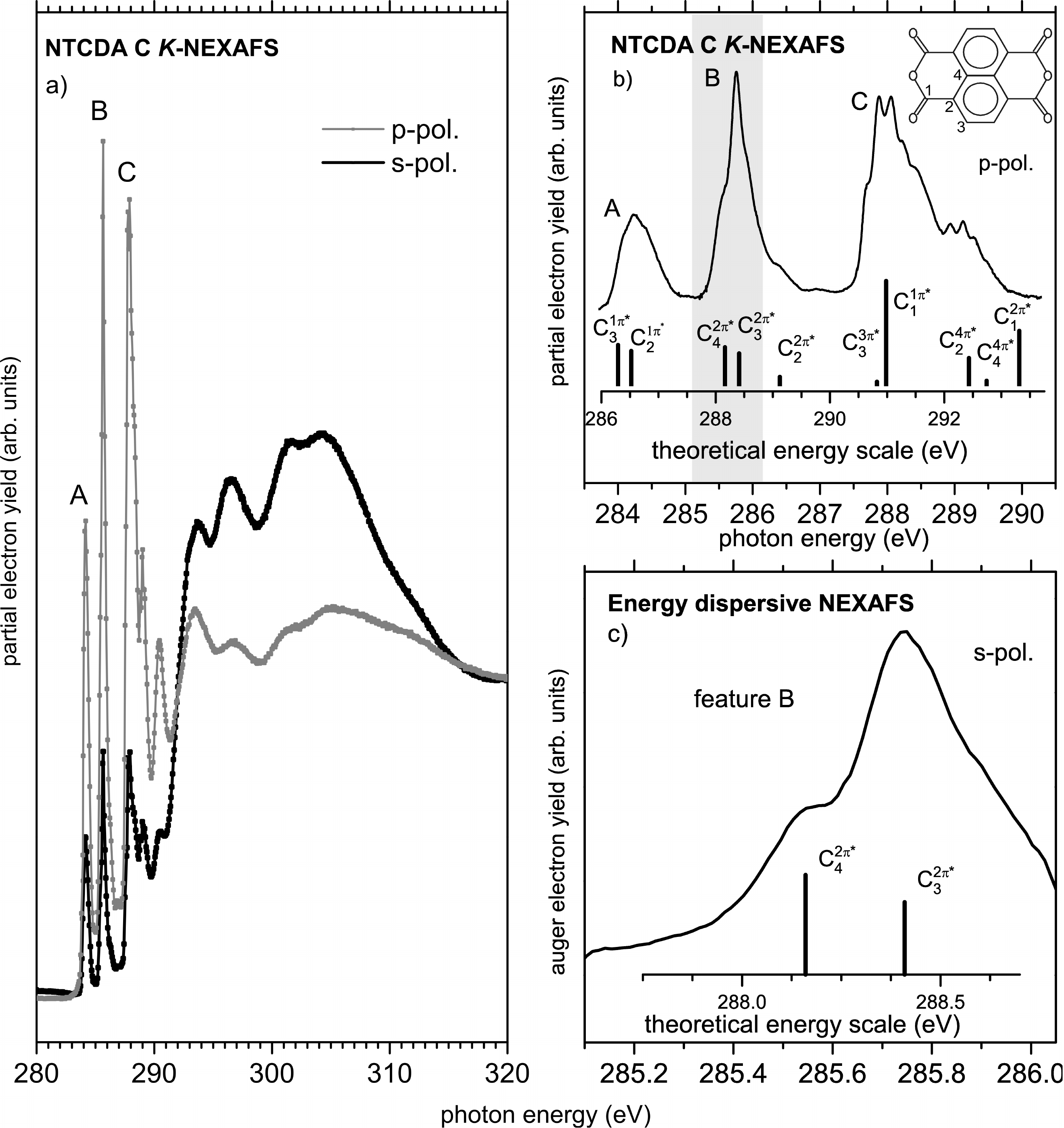}
	\caption{(a) C K-NEXAFS spectra of a NTCDA multilayer film as prepared at 95~K, recorded with \textit{s} (black line) and \textit{p} polarization (gray line) of the incident beam. (b) Close-up of the $\pi^*$-resonances (denoted as A--C) recorded with \textit{p} polarization, compared to the result of a Hartree-Fock calculation with the GSCF3 code \cite{kosugi_strategies_1987}. The theoretical energy scale was shifted and scaled to fit the experimental data. The calculated electronic transitions are marked according to the respective carbon 1s sites C1--C4 (see inlay) and final state orbitals $\pi^*$, 2$\pi^*$, 3$\pi^*$. The shaded region indicates the energy window recorded in the energy dispersive mode, which is displayed in (c).}
	\label{fig:fig1}
\end{figure}

\indent For the model system NTCDA [see molecular structure in Fig. \ref{fig:fig1}(b)] on a Ag(111) surface, different structural phases exist for multilayer films \cite{gador_manipulation_1998, scholl_electron-vibron_2004, braatz_vibrational_2012}. Growth at substrate temperatures below 100~K results in morphologically smooth films with preferentially flat lying molecules and without long-range order \cite{gador_manipulation_1998}. Heating the samples to temperatures above 180~K leads to an increasing crystallinity in the films and changes the molecular orientation to a more upright standing configuration. During this transition a transient phase occurs, which exists for only about 10~min and which shows the characteristic changes of the signature of the NEXAFS spectra of particular interest for this work. The fast acquisition of x-ray absorption data is thus crucial and the experiments were therefore performed at the beam line UE52-PGM of BESSY II in Berlin Ref. \cite{note2}. The beam line was operated in the energy dispersive mode, where the beam is focused on the sample without an exit slit, thus leading to a spatial dispersion of photon energies on the sample in vertical direction \cite{batchelor_energy-dispersive_2007}. By use of an electron spectrometer with adapted spatial resolution (VG-SCIENTA R4000), NEXAFS spectra can be recorded in a multichanneling mode with a possible acquisition time of less than 1~sec and an energy resolution better than 50~meV at the C~\textit{K}-edge. \\


\indent Figure \ref{fig:fig1}(a) shows the C \textit{K}-NEXAFS spectra of a four monolayer (4~ML) thick NTCDA film which was prepared and measured at a substrate temperature of 95~K \cite{note2}. The absorption signal of the $\pi^*$-resonances between 282~eV and 290~eV shows a characteristic dependence on the polarization of the incident light, which allows determining the average molecular orientation angle with respect to the substrate surface \cite{stohr_nexafs_1992}. The strong dichroism in Fig. \ref{fig:fig1}(a) indicates a preferential flat lying orientation of the NTCDA molecules in the film, consistent with previous investigations \cite{gador_manipulation_1998, scholl_electron-vibron_2004}. Figure \ref{fig:fig1}(b) shows the $\pi^*$ resonances of the \textit{p} polarized spectrum on an expanded energy scale. According to Hartree-Fock calculations \cite{kosugi_strategies_1987} plotted on the bottom of  Figure \ref{fig:fig1}b), the spectrum can be explained by the superposition of several electronic transitions from the four symmetry-inequivalent carbon 1s sites C1--C4 [see inset in Fig. \ref{fig:fig1}(b)] to the final state orbitals 1$\pi^*$, 2$\pi^*$, 3$\pi^*$. In addition, the electronic transitions show a characteristic coupling to vibrations, leading to the rich fine structure of the three main signals, denoted by A, B, and C in Fig. \ref{fig:fig1}(b) \cite{scholl_electron-vibron_2004}. While the spectra in Fig. \ref{fig:fig1}(a) and 1(b) were recorded by scanning the photon energy in the normal beam line mode with an acquisition time of about 30~min per spectrum,  Fig. \ref{fig:fig1}(c) shows an energy dispersive NEXAFS experiment of the energy regime of feature B [shaded region in Fig. \ref{fig:fig1}(b)]. The integration time was 30~sec and the data are in very good agreement with feature B of Fig. \ref{fig:fig1}(b). The fine structure is well resolved, and according to the calculations in Fig. \ref{fig:fig1}(b) the signal in the photon energy range between 285.1 and 286.0~eV corresponds to the two electronic transitions C4 1s-2$\pi^*$ and C3 1s-2$\pi^*$, which are originating from carbon atoms located on the naphthalene core [see Fig. \ref{fig:fig1}(b)]. As demonstrated in Ref. \cite{scholl_electron-vibron_2004}, these electronic transitions lead to the excitation of vibrations of the aromatic core with vibrational energies of around 90~meV. 

\begin{figure}
	\centering
			\includegraphics[width=0.50\textwidth]{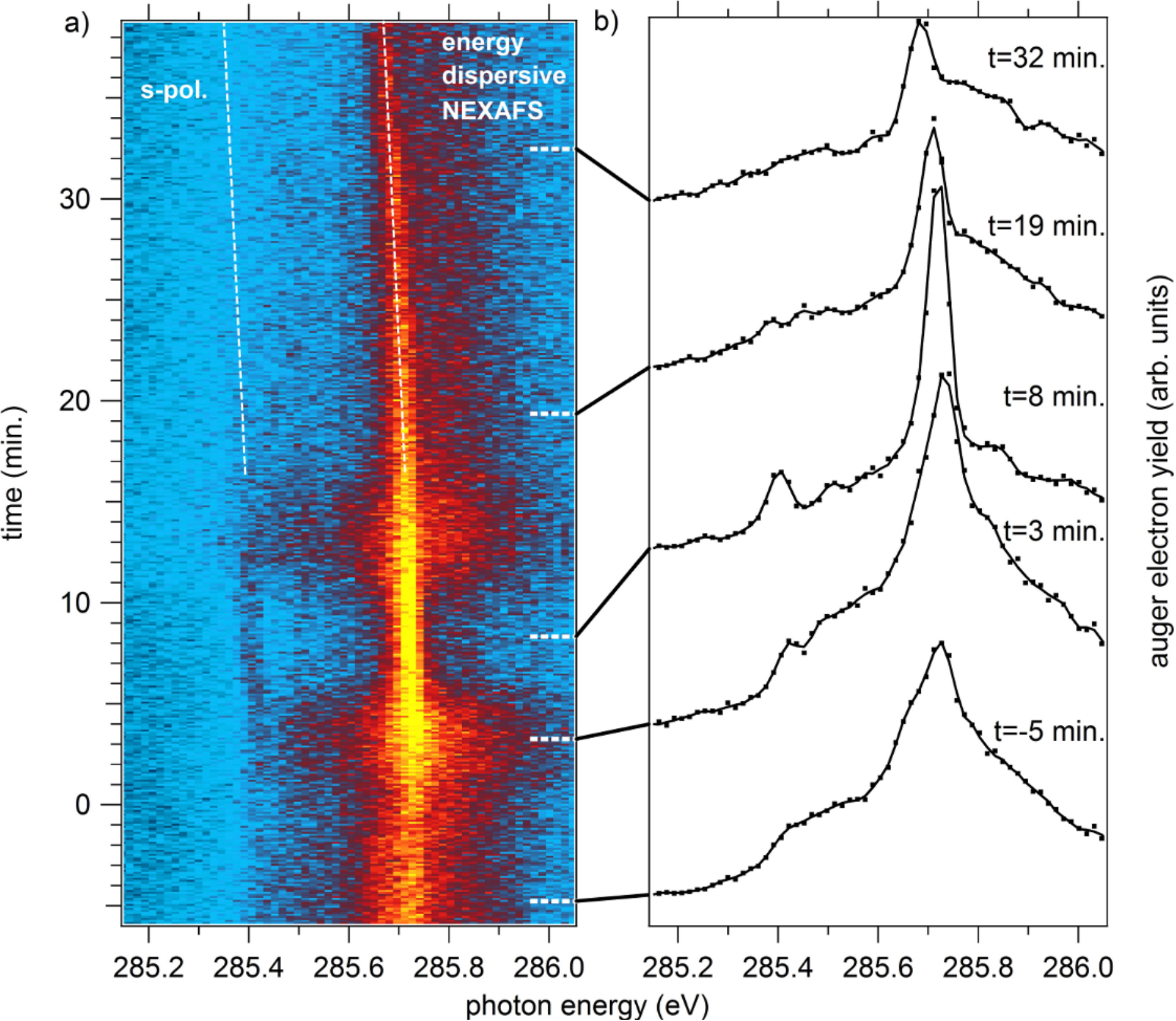}
	\caption{(a): 2D color map of the energy dispersive NEXAFS spectra of feature \textit{B} recorded with \textit{s} polarized light plotted against time. Yellow refers to high, blue to low intensity. The 4~ML thick film was prepared at 95~K and heated to 220~K within 1~min. \textit{t}=0 refers to the time when a constant sample temperature of 220~K was reached. The NEXAFS data was collected continuously during annealing for about 45~min. The vertical  white dashed lines are guidelines for the eye to illustrate the energy shift of the main components. (b) NEXAFS spectra of feature B derived by integration of the  intensity map of (a) over 3~min at the times indicated by the horizontal white dashed lines (black dots: original data; solid line: 3-point average).}
	\label{fig:fig2}
\end{figure}

\indent The color plot in Fig. \ref{fig:fig2}(a) shows the evolution of feature B with time during an energy dispersive NEXAFS experiment. The sample (thickness about 4~ML) was prepared at 95~K and then quickly (in about 6~min) heated to 220~K in order to induce the transition from flat lying to upright standing molecules. On the y axis \textit{t}=0 refers to the time when a constant sample temperature of 220~K was reached. On the bottom of Fig. \ref{fig:fig2}(a), i.e., directly after preparation at 95~K, the spectra resemble the typical absorption signal for films prepared at low temperature [see Fig. \ref{fig:fig1}(c)]. This is demonstrated by Fig. \ref{fig:fig2}(b), which displays spectra derived after integration of the intensity map in Fig. \ref{fig:fig2}(a) over  intervals of 3~min at the times indicated by the white horizontal lines. 

Shortly after a substrate temperature of 220~K is reached, the line shape of feature B is significantly altered. Starting at \textit{t}=2~min the intensity of the absorption peaks at 285.4~eV and 285.7~eV increases, while it decreases in between. In addition, the line width of the most prominent signals decreases, which can be observed best in the spectrum in Fig. \ref{fig:fig2}(b) extracted after 8~min at 220~K. Moreover, an energy shift of the main spectral components towards lower energy occurs, which is indicated by the white dashed lines in Fig. \ref{fig:fig2}(a). The energy shift sums up to about $\sim$ 80~meV after 35~min. While the overall intensity increases until about \textit{t}=12~min, the absorption signal is constantly decreasing after the sample was kept at 220~K for more than 12~min. \\

\begin{figure}
	\centering
				\includegraphics[width=0.50\textwidth]{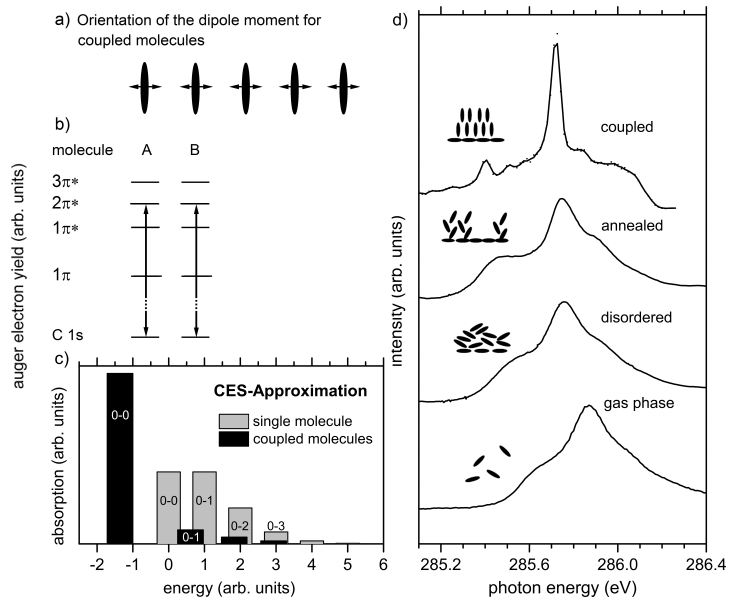}
	\caption{(a) Illustration of the coupling of NTCDA molecules in the transient phase. The transition dipole moments are oriented perpendicular to the molecular plane. (b) Schematic energy level diagram of adjacent NTCDA molecules after resonant core excitation. See text for details. (c) Result of a coherent exciton scattering calculation: The adiabatic component of the electronic excitation (0-0) and the corresponding vibronic progression (0--1, 0--2, 0--3) are plotted for isolated (gray) and coupled (black) molecules. (d) NEXAFS spectra of  NTCDA feature B in the gas phase \cite{holch_new_2011}, in the disordered condensed phase \cite{scholl_electron-vibron_2004}, in the ordered condensed phase after annealing \cite{scholl_electron-vibron_2004}, and in the transient phase (from bottom). The spectra were normalized arbitrarily for better comparison of the line shape.}
	\label{fig:fig3}
\end{figure}


\indent This variation of general intensity can be explained by the change of orientation angle of the NTCDA molecules from flat lying to a more upright configuration, which occurs after the sample temperature is increased to 220~K \cite{gador_manipulation_1998}. This leads to an immediate intensity increase in the C \textit{K}-NEAXFS data recorded with \textit{s} polarization \cite{stohr_nexafs_1992}. However, this intensity increase is counteracted by a change of film morphology, since the increased sample temperature also leads to a roughening of the originally closed film and to formation of 3D islands. Because of the limited probing depth of about 1~nm \cite{graber_experimental_2011} of the Auger electrons mainly contributing to the NEXAFS signal in electron yield mode, this second process leads to an overall reduction of the absorption signal. Moreover, the morphology change includes material transport and occurs on a slower time scale than the change of orientation angle, thus accounting for the observed delayed intensity decrease \cite{Scholz_2012}.  

\indent We will in the following concentrate on the particular changes of the spectral signature of the NTCDA C~K-NEXAFS which occur during the transition. Figure \ref{fig:fig3}(d) compares the NEXAFS spectra of feature B for four different NTCDA samples. On the bottom gas phase data are displayed, which is representative of the isolated molecule \cite{holch_new_2011}. Film growth at low sample temperature (\textit{T}=95~K) results in the disordered phase with preferentially flat lying molecules (second spectrum from bottom) \cite{scholl_electron-vibron_2004}, while annealing to temperatures above 180~K results in the ordered phase with more upright molecular orientation (third spectrum from bottom) \cite{scholl_electron-vibron_2004}. The data of the transient phase during the transition are plotted on top. 
The line shape in the disordered phase resembles the gas phase, while the entire spectrum is redshifted by about 100~meV upon condensation. The ordering of the NTCDA molecules upon annealing leads to a slight decrease of the line width and to a more pronounced shoulder on the low energy side of feature B. This trend is followed if the transient phase is examined. The most intense peak at 285.7~eV is now very sharp and the low energy shoulder has developed into a well distinguishable peak at 285.4~eV. Moreover, additional peaks can be resolved at 285.5~and 285.6~eV as well as at 285.8~eV. The energetic separation of these signals from the prominent peaks at 285.4~and 285.7~eV corresponds well to the vibronic energy of the NTCDA modes coupled to the C4 1s-2$\pi^*$ and C3 1s-2$\pi^*$- transitions, respectively \cite{scholl_electron-vibron_2004}. We thus straightforwardly interpret the spectrum of feature B in the transient phase as due to the prominent adiabatic transitions C4 1s-2$\pi^*$ at 285.4~eV and C3 1s-2$\pi^*$ at 285.7~eV with adjacent vibronic states. The main difference from the other condensed and gas phase spectra lies in a reduced line width of the involved signals and in a change of the vibronic envelope, which obviously shifts intensity towards the adiabatic transition in the transient phase.
 
\indent An explanation for this observation can be given on the basis of a coupling of the transition dipole moments of adjacent molecules. 
The coupling strength strongly depends on the arrangement of the molecules and on the symmetry of the respective transition moments. In the dipole approximation the coupling strength $V_{nm}$ between molecules n and m can be described by 
\begin{equation}
V_{nm}=\left|\vec{\mu}_n\right|\left|\vec{\mu}_m\right|\cdot\left(\frac{\vec{n}_{\vec{\mu}_n}\vec{n}_{\vec{\mu}_m}}{R^3_{nm}}-\frac{3(\vec{n}_{\vec{\mu}_n}\vec{r}_{nm})(\vec{n}_{\vec{\mu}_m}\vec{r}
_{nm})}{R^5_{nm}}\right), 
\end{equation}
were $\vec{\mu}_n$ and $\vec{\mu}_m$ are the transition dipole moments on molecules n and m, respectively, which are separated by $\vec{r}$. For a C~1s-2$\pi^*$ transition with a dipole moment perpendicular to the molecular plane, the coupling is thus largest for a head-to-tail arrangement of the dipoles (commonly referred to  as J aggregates); i.e., a configuration as sketched in Fig. \ref{fig:fig3}(a). \\
\indent The proposed dipole-dipole coupling mechanism is illustrated in Fig. \ref{fig:fig3}(b). The C~1s-2$\pi^*$ excitation of molecule A couples to the same transition on the neighboring molecule B. \\
\indent The changes of the vibrational envelope observed in our NEXAFS data can be understood in the framework of the coherent exciton scattering (CES) theory \cite{briggs_sum_1970, eisfeld_j-band_2002, eisfeld_j-_2006, walczak_exchange_2008}. This one-particle Green's function approach obeys Dyson's equation and describes the excitation of coherently delocalized quasiparticles. In other words, this means that the hole-electron pair propagates to the adjacent molecule fast compared to the molecular vibration frequencies. Therefore, the originally excited molecule remains preferentially in its vibrational ground state. As a result, intensity is shifted towards the adiabatic transition in the vibrational envelope observed in experiment. This is illustrated by Fig. \ref{fig:fig3}(c), which shows the result of a CES calculation of the vibronic progression of a single molecule and of coupled molecules in a configuration as sketched in Fig. \ref{fig:fig3}(a) \cite{note}. The intensity ratio of the adiabatic transition (0-0) to the higher vibrational states (0--1, 0--2, and 0--3) changes strongly, and in the case of sufficiently strong coupling the oscillator strength accumulates in the 0-0~transition. Moreover, a marked energy shift of the entire vibronic progression towards lower energy can be observed in the case of the coupled molecules, which also matches our experimental observation qualitatively [see Fig. \ref{fig:fig2}(a)].

\indent In the transient phase we also obverse a reduced line width of the C 1s-2$\pi^*$ transitions, as evidently seen in Fig. \ref{fig:fig3}(d). 
For a quantitative analysis the FWHM of the gas phase and of the transient phase spectra was estimated to 211~($\pm$10)~meV and 74~($\pm$10)~meV respectively, for the most prominent signals in the data of Fig. \ref{fig:fig3}(d). According to Knapp et al. \cite{knapp_lineshapes_1984} the decrease of the FWHM $\Delta$ of the absorption signal with the number of coherently coupled molecules $N$ can be described by the equation: 
\begin{equation}
\Delta\propto\frac{1}{\sqrt{N}}.
\end{equation} 
With the FWHM derived from our NEXAFS data of isolated and coherently coupled molecules, i.e. $\Delta_{isol.}$ and $\Delta_{coup.}$, respectively, a number of about $N$=8$\pm$3 can be estimated in the transient phase.\\

\indent In conclusion, we have reported on an x-ray absorption experiment which demonstrates the importance of a correct consideration of the core hole-electron correlation for condensates of organic molecules. During the transition from a disordered to an ordered film arrangement, a transient phase occurs for NTCDA/Ag(111) which is characterized by a distinctly different spectral signature. The spectra of this transient phase exhibit a reduced line width, a redshift, as well as a substantial change of the vibronic profile. All these experimental observations can be explained by the coupling of the transition dipole moments of adjacent molecules, a model, which is well established for optical spectroscopies. This result has two immediate implications: Firstly, it shows that the excitation delocalizes over several molecules (a number of about 10 can be determined from our data). This comprises the existence of a delocalized core hole, which is involved in the excitation. This implication is not understood and needs further consideration. Note that other models have been considered but could not explain our experimental observation.  

Secondly, the very accurate explanation of our experimental data by this model based on the coupling of transition dipoles strongly suggests that the oscillator strength of the transition dipole moments involved in the core-to-valence excitations is substantially larger [a value of about 0.37~a.u. can be estimated from the experimental data and eq. (1)] than derived from calculations on isolated molecules (about 0.04~a.u. for the respective NTCDA transitions \cite{scholl_electron-vibron_2004}). This discrepancy may be explained by the additional contribution of the wave function overlap between neighboring molecules \cite{fuckel_theoretical_2008,curutchet_does_2008,yamagata_nature_2011, yamagata_designing_2012}. For close-packed $\pi$-conjugated molecules in a face-to-face arrangement, neighboring dyes can have a considerable overlap of the frontier orbital wave functions \cite{kera_very_2002, koch_evidence_2006, ueno_electron_2008, machida_highest-occupied-molecular-orbital_2010}. This can lead to increased intermolecular charge transfer transition dipole moments \cite{bredas_charge-transfer_2004, bredas_molecular_2009}. This so-called short-range component of the coupling becomes significant, e.g., for naphthalene dimers already at a separation of about 6~\AA~\cite{scholes_electronic_1994} and can thus be expected to be substantial in the present case, where the intermolecular distance between the NTCDA molecules is only around 3~\AA.~Moreover, the relatively small coupling values for core-to-valence excitations predicted by theory are usually derived from a point-dipole approach, which may find its restrictions in the present case of a very dense molecular packing, where the intermolecular distance of around 3~\AA~is much smaller than the size of the NTCDA molecules ($\sim$9~\AA~along the long axis). Further efforts to develop a comprehensive theoretical description of the core excitation in molecular condensates are thus highly desirable.

\begin{acknowledgments}
We thank Patrick Hoffmann (BESSY), David Batchelor (KIT) and the BESSY staff for assistance during beam time. We acknowledge stimulating discussions with Alexander Eisfeld and Peter Jakob. This work was financially supported by the Deutsche Forschungsgemeinschaft (Grants No. FOR1162, No. GRK1221, and No. RE1469/9-1) and the Bundesministerium f\"ur Bildung und Forschung (Grant No. 03SF0356B).  \\
\end{acknowledgments}



\begin{thebibliography}{36}
\bibitem [{not()}]{note}%
  \BibitemOpen
  \href@noop {} {\bibinfo  {journal} {Note that the coupling strength was
  chosen arbitrarily in order to match the observed changes of the vibronic
  progression}\ }\BibitemShut {NoStop}%
\end{thebibliography}
%

\end{document}